\documentclass[twoside,twocolumn,9pt]{article}
\usepackage{extsizes}
\usepackage[super,sort&compress,comma]{natbib} 
\usepackage[version=3]{mhchem}
\usepackage[left=1.5cm, right=1.5cm, top=1.785cm, bottom=2.0cm]{geometry}
\usepackage{balance}
\usepackage{times,mathptmx}
\usepackage{sectsty}
\usepackage{graphicx} 
\usepackage{lastpage}
\usepackage[format=plain,justification=justified,singlelinecheck=false,font={stretch=1.125,small,sf},labelfont=bf,labelsep=space]{caption}
\usepackage{float}
\usepackage{fancyhdr}
\usepackage{fnpos}
\usepackage[english]{babel}
\addto{\captionsenglish}{%
  \renewcommand{\refname}{Notes and references}
}
\usepackage{array}
\usepackage{droidsans}
\usepackage{charter}
\usepackage[T1]{fontenc}
\usepackage[usenames,dvipsnames]{xcolor}
\usepackage{setspace}
\usepackage[compact]{titlesec}
\usepackage{hyperref}

\usepackage{epstopdf}

\definecolor{cream}{RGB}{222,217,201}

\begin{document}

\pagestyle{fancy}
\thispagestyle{plain}


\makeFNbottom
\makeatletter
\renewcommand\LARGE{\@setfontsize\LARGE{15pt}{17}}
\renewcommand\Large{\@setfontsize\Large{12pt}{14}}
\renewcommand\large{\@setfontsize\large{10pt}{12}}
\renewcommand\footnotesize{\@setfontsize\footnotesize{7pt}{10}}
\renewcommand\scriptsize{\@setfontsize\scriptsize{7pt}{7}}
\makeatother

\renewcommand{\thefootnote}{\fnsymbol{footnote}}
\renewcommand\footnoterule{\vspace*{1pt}%
\color{cream}\hrule width 3.5in height 0.4pt \color{black} \vspace*{5pt}} 
\setcounter{secnumdepth}{5}

\makeatletter 
\renewcommand\@biblabel[1]{#1}            
\renewcommand\@makefntext[1]%
{\noindent\makebox[0pt][r]{\@thefnmark\,}#1}
\makeatother 
\renewcommand{\figurename}{\small{Fig.}~}
\sectionfont{\sffamily\Large}
\subsectionfont{\normalsize}
\subsubsectionfont{\bf}
\setstretch{1.125} 
\setlength{\skip\footins}{0.8cm}
\setlength{\footnotesep}{0.25cm}
\setlength{\jot}{10pt}
\titlespacing*{\section}{0pt}{4pt}{4pt}
\titlespacing*{\subsection}{0pt}{15pt}{1pt}

\fancyfoot{}
\fancyfoot[RO]{\footnotesize{\sffamily{1--\pageref{LastPage} ~\textbar  \hspace{2pt}\thepage}}}
\fancyfoot[LE]{\footnotesize{\sffamily{\thepage~\textbar\hspace{3.45cm} 1--\pageref{LastPage}}}}
\fancyhead{}
\renewcommand{\headrulewidth}{0pt} 
\renewcommand{\footrulewidth}{0pt}
\setlength{\arrayrulewidth}{1pt}
\setlength{\columnsep}{6.5mm}
\setlength\bibsep{1pt}

\makeatletter 
\newlength{\figrulesep} 
\setlength{\figrulesep}{0.5\textfloatsep} 

\newcommand{\topfigrule}{\vspace*{-1pt}%
\noindent{\color{cream}\rule[-\figrulesep]{\columnwidth}{1.5pt}} }

\newcommand{\botfigrule}{\vspace*{-2pt}%
\noindent{\color{cream}\rule[\figrulesep]{\columnwidth}{1.5pt}} }

\newcommand{\dblfigrule}{\vspace*{-1pt}%
\noindent{\color{cream}\rule[-\figrulesep]{\textwidth}{1.5pt}} }

\makeatother

\twocolumn[
  \begin{@twocolumnfalse}
\vspace{3cm}
\sffamily
\begin{tabular}{m{18cm} p{0cm} }

\noindent\LARGE{\textbf{Density functional theory calculation and thermodynamic analysis of the bridgmanite surface structure}} \\
 & \vspace{0.3cm} \\

 \noindent\large{Ming Geng$^{\ast}$\textit{$^{ a,}$}\textit{$^{b}$} and Hannes J\'onsson\textit{$^{ c,}$}\textit{$^{d}$}} \\

\end{tabular}

 \end{@twocolumnfalse} \vspace{0.6cm}

  ]

\renewcommand*\rmdefault{bch}\normalfont\upshape
\rmfamily
\section*{}
\vspace{-1cm}

\footnotetext{\textit{$^{\ast}$~gengming@mail.iggcas.ac.cn}}
\footnotetext{\textit{$^{a}$~Key Laboratory of Earth and Planetary Physics, Institute of Geology and Geophysics, Chinese Academy of Sciences, Beijing 100029, China.}}
\footnotetext{\textit{$^{b}$~Institutions of Earth Science, Chinese Academy of Sciences, China}}
\footnotetext{\textit{$^{c}$~Faculty of Physical Sciences, University of Iceland, 107 Reykjav\'{i}k, Iceland}}
\footnotetext{\textit{$^{d}$~Dept. of Energy Conversion and Storage, Technical University of Denmark, DK-2800 Kgs. Lyngby, Denmark}}

 \footnotetext{\dag~Electronic Supplementary Information (ESI) available: [details of any supplementary information available should be included here]. }




\sffamily{\textbf{Bridgmanite, a high temperature and pressure form of \ce{MgSiO$_3$}, is believed to be Earth's most abundant mineral and responsible for the observed seismic anisotropy in the mantle. Little is known about surfaces of bridgmanite but knowledge of the most stable surface terminations is important for understanding various geochemical processes as well as likely slip planes. A density functional theory based thermodynamic approach is used here to establish the range of stability of bridgmanite as well as possible termination structures of the (001), (010), (100) and (011) surfaces as a function of the chemical potential of oxygen and magnesium. The results presented provide a basis for further theoretical studies of the chemical processes on bridgmanite surfaces in the Earth's mantle and slip plane analysis.}}\\


\rmfamily 



Bridgmanite is believed to be the most abundant rock-forming mineral in the Earth, making up 38\% of the Earth's interior volume\cite{RN298344}. 
Its importance in the mantle has long been recognized\cite{RN298337}. The composition is \ce{MgSiO3} in an orthorhombic \ce{ABO3} perovskite structure.
and it had long been called magnesium perovskite before it was officially named after the 1946 Nobel laureate in physics Percy W. Bridgman \cite{RN298343, RN298338}. Magnesium atoms are is in the A sites and silicon atoms in the B sites. 
Numerous experimental\cite{RN298346,RN298350,RN298349} and computational \cite{RN298372,RN298352,RN298354} investigations of the crystal have been reported. Bridgmanite is believed to be stable from 660 km nearly down to the core-mantle boundary region at a depth of 2900km\cite{RN298348,RN298347, RN298351}. 
An observed seismic shear wave anisotropy in the Earth's upper mantle has been ascribed to a preferred orientation of induced deformation of bridgmanite is the most plausible explanation to observations of the Earth's uppermost mantle seismic shear wave anisotropy 
\cite{RN298354}. 
Deformation experiments have been carried out on  bridgmanite\cite{RN298373,RN298374}
as well as theoretical atomic level simulations\cite{RN298271} based on the Peierls-Nabarro model \cite{RN298370,RN298371} and NEB method.\cite{RN2626} The 
rheology of bridgmanite is characterized by high lattice friction opposed to dislocation glide 
which requires high stress levels 
and this correlates well with the viscosity jump associated with the uppermost lower mantle \cite{RN298272}. 
A more complete understanding of the rheological properties of bridgmanite require further studies of 
polycrystalline aggregates of bridgmanite and grain boundaries.

While several experiments and calculations have been carried out on the properties of the bridgmanite crystal, 
little work has been done to characterize the surface properties. The only work we are aware of are theoretical calculations of
Alfredsson et al.\cite{RN298264} who used effective pair potential functions tested and parametrized by plane wave density functional theory (DFT) calculations. 
They found the
(001) surface to be the most stable surface of bridgmanite by calculating the cleavage energy of various surfaces.

The MgO- and SiO2-termination were compared only for the (001) surface. 
It is important to determine the structure and stability of the various bridgmanite surfaces and surface terminations for a range of environmental condition in order to 
establish the surface reactivity towards volatile compounds such as \ce{H2O} and \ce{CO2}, and the effect of surface impurities.

In the present article, results of DFT and thermodynamic calculations the stability of bridgmanite with respect to other solids as a function of temperature and presssure, as well as the structure and stability of various bridgmanite surface terminations\footnote{The termination structures can be found in the ESI. } 
are presented for a range in oxygen and magnesium chemical potential.
This work lays the foundations for further modeling of the rheology of bridgmanite and surface chemisitry.


\begin{figure}[h]
  \centering
    \includegraphics[width=0.98\linewidth]{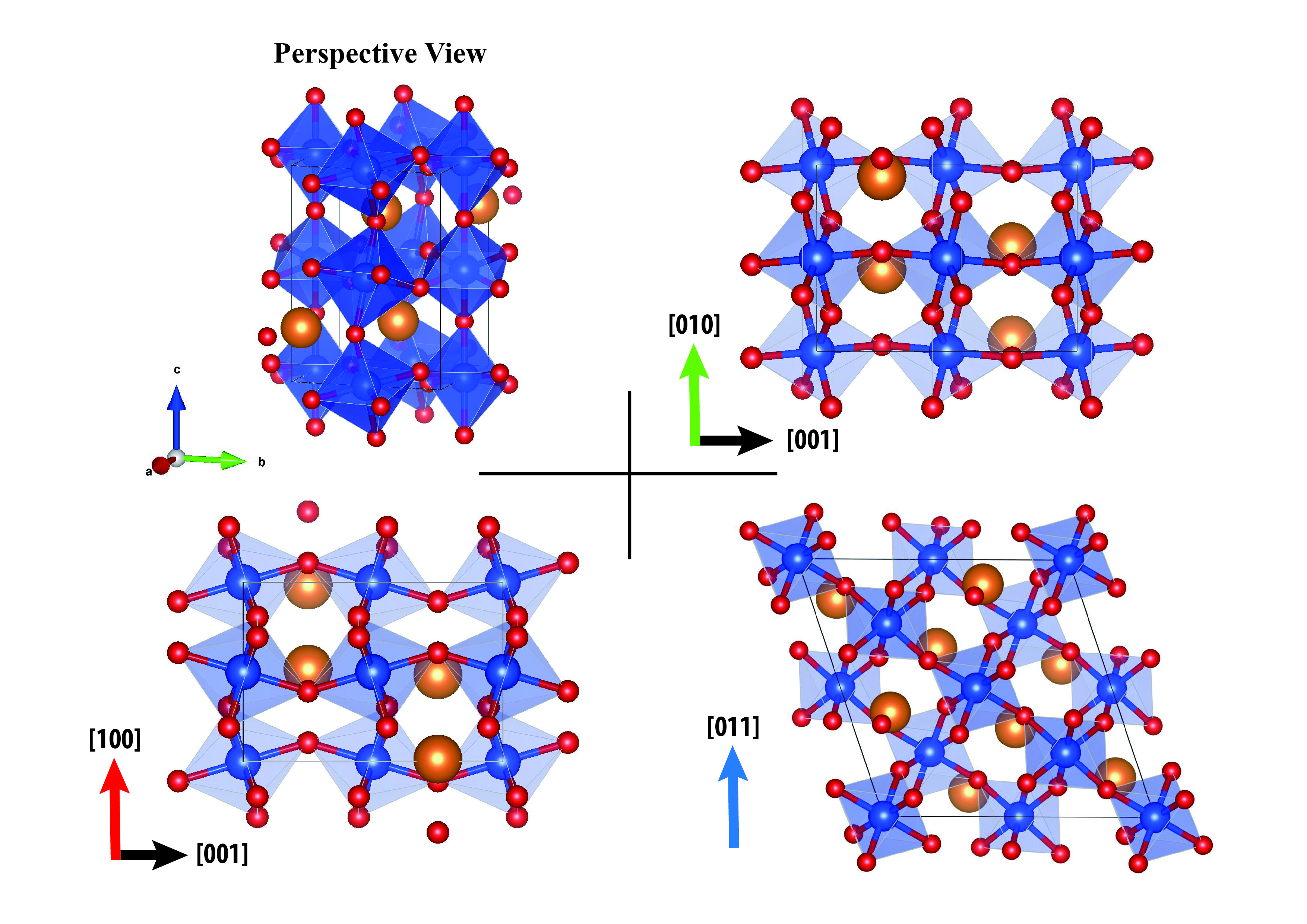}
      \caption{Structure of the bridgmanite unit cell with side views along various directions.}
    \label{fig:UnitCell}
  \end{figure}

  Unlike the more common cubic perovskite structures, the neighboring $ SiO_6 $ octahedral in bridgmanite are rotated with respect to the [001] axis and form a \textit{Pbnm} orthorhombic structure as shown in Fig \ref{fig:UnitCell}. The (001) surface is non-polar while (011) is a polar surface. 
  Surfaces with the various terminations, including also (010) and (100) surfaces, are created by cleavage of the crystal along the corresponding axis. 
  Cleaving the crystal will creates two complementary surfaces and the cleavage energy per a unit cell is calculated as 

\begin{equation}
E_{cleave}(i, j) = - \frac{1}{2}[E^i_{slab}+E^j_{slab}-2E^{bulk}_{MgSiO_3}]
\end{equation}

Cleavage of the crystal along the [001] direction can be used to create two pairs of complementary surfaces. 
For the [010] and [100] directions there are 4 pairs, while for the [011] direction there are 5 pairs of complementary surfaces,
as listed in table \ref{tbl:cleavage}.

\begin{table*}[]
  \small
    \caption{\ Cleavage energies per unitcell ($J\//m^2$) of bridgmanite in different directions}
    \label{tbl:cleavage}
    \begin{tabular*}{\textwidth}{@{\extracolsep{\fill}}l|l|l|l|l|l|l|l}
      \hline
      \multicolumn{2}{c|}{(001)}&\multicolumn{2}{c|}{(010)}&\multicolumn{2}{c|}{(100)}&\multicolumn{2}{c}{(011)}\\
      \hline
      MgO:SiO2&2.43&SiO:Mg&2.21&Mg:Si2OMg&3.76&Si2:MgO3&1.30\\
      O:SiO   &6.55&Si2O:MgO&1.30&MgO2:MgO2&2.22&O:MgO2-MgO3&1.06\\
              &    &O2:O2&2.08&O4:SiO&0.83&MgO2-O2:MgO-O&2.67 \\
              &    &O4:MgSiO&1.89&O5:Si2OMg2&2.85&MgO-O2:MgO-O2&1.88\\
              &    &&&            &    &MgO:MgO3-O3&1.13 \\
      \hline
    \end{tabular*}
  \end{table*}

The energetics are calculated using DFT implemented with the projector-augmented wave (PAW) method\cite{RN298234} and Perdew-Burke-Ernzerhof (PBE) functional approximation\cite{RN298236}. A plane wave basis set with a 600 eV kinetic energy cutoff was used and a Monkhorst-Pack k-point mesh of 8$\times$8$\times$8 for bulk phases and 8$\times$8$\times$1 for slab calculations of surface terminations.   In order to minimize the influence caused by the periodic boundary condition, a 30  \AA \space thick vacuum region was included between   periodic images of the slab. The VASP software was used in the calculations\cite{RN298230,RN298231,RN298232,RN298233}. 

Vibrational contribution to the Gibbs free energy were evaluated using the phonon density of states calculated using the density functional perturbation theory as implemented in the VASP/Phonopy software\cite{RN2771}.

The cleavage energy is the energy required to split a crystal into two parts with complementary terminations. The cleavage energy is related to the energy of the two surfaces formed. The Gibbs surface free energy is the excess energy of a semi-infinite crystal in contact with chemical reservoirs and can be used to analyze the stability of the various surface terminations. 
Slabs with identical terminations on both sides are used to evaluate the Gibbs free energy. The most stable surface is the one with the smallest surface Gibbs free energy for given chemical potentials of the various atomic species. The deviation of chemical potentials from their reference state ($\Delta \mu = \mu - G^{reference}_{state} $) are used as as variables 

For bridgmanite this is
\begin{equation}
  \begin{aligned}
  \Omega_i &=\dfrac{1}{2A}(G^i_{slab}-N_{Mg}\mu_{Mg}-N_{Si}\mu_{Si}-N_O\mu_{O})\\& =\phi^i-\dfrac{1}{2A}(N^i_{Mg}-N^i_{Si}\dfrac{N^{bulk}_{Mg}}{N^{bulk}_{Si}})\Delta\mu_{Mg}-\dfrac{1}{2A}(N^i_{O}-N^i_{Si}\dfrac{N^{bulk}_{O}}{N^{bulk}_{Si}})\Delta\mu_O
  \end{aligned}
\end{equation}
where 
\begin{equation}
  \begin{aligned}
    \phi ^i&=\frac{1}{2A}(G^i_{slab}-N^i_{Si}G^{bulk}_{MgSiO_3})-\dfrac{1}{2A}(N^i_{Mg}-N^i_{Si}\dfrac{N^{bulk}_{Mg}}{N^{bulk}_{Si}})G^{bulk}_{Mg}\\&-\dfrac{1}{4A}(N^i_{O}-N^i_{Si}\dfrac{N^{bulk}_{O}}{N^{bulk}_{Si}})E_{O_2}
  \end{aligned}
\end{equation}
$\Omega_i$ is the surface Gibbs free energy per unit area, $G^i$ is the Gibbs free energy of crystals, $N^i $ is the number of atoms of type $i$ in crystal or surface; and $A$ is the surface area of the slab termination.
In order for the \ce{MgSiO3} crystal to be stable with respect to segregation of Mg, Si or \ce{O2} molecules 
the following conditions on $\Delta \mu$ need to be satisfied
\begin{equation}\Delta \mu_{Mg} \le 0\end{equation}
\begin{equation}\Delta \mu_{Mg} + 3 \Delta \mu_{O} \ge \Delta g_f(MgSiO_3)\end{equation}
$\Delta g_f$ is the Gibbs free energy of formation of crystal. Similarly, precipitation of \ce{MgO} and \ce{SiO2} do not occur if:
\begin{equation}\Delta \mu_{Mg} +\Delta \mu_{O} \le \Delta g_f(MgO)  \end{equation}
\begin{equation}\Delta \mu_{Mg} +\Delta \mu_{O} \ge \Delta g_f(MgSiO_3)-\Delta g_f(SiO_2)\end{equation}

The bridgmanite crystal is thermodynamically stable when all these conditions are met. Previous studies have shown that silicate perovskite type minerals can be stable at high pressure\cite{RN298408}. As pressure is increased, the crystal structure can change while the chemical composition remains the same. In order to determine the region of stability area of bridgmanite in a $\Delta \mu$ graph, it is important to compare with the right crystal structures of the various segregation products.
  
MgO will not change its crystal structure from B1 to B2 phase below 300GPa\cite{RN298421}, but \ce{SiO2} will show a series of phase transition when pressure is increased\cite{RN298423}. At ambient pressure, \ce{SiO2}  is quartz, but stishovite becomes more stable at high pressure. Especially in the lower mantle condition where bridgmanite is stable, stishovite will be the predominant form of \ce{SiO2}.




\begin{figure}[h]
 \centering
 \includegraphics[width=0.98\linewidth]{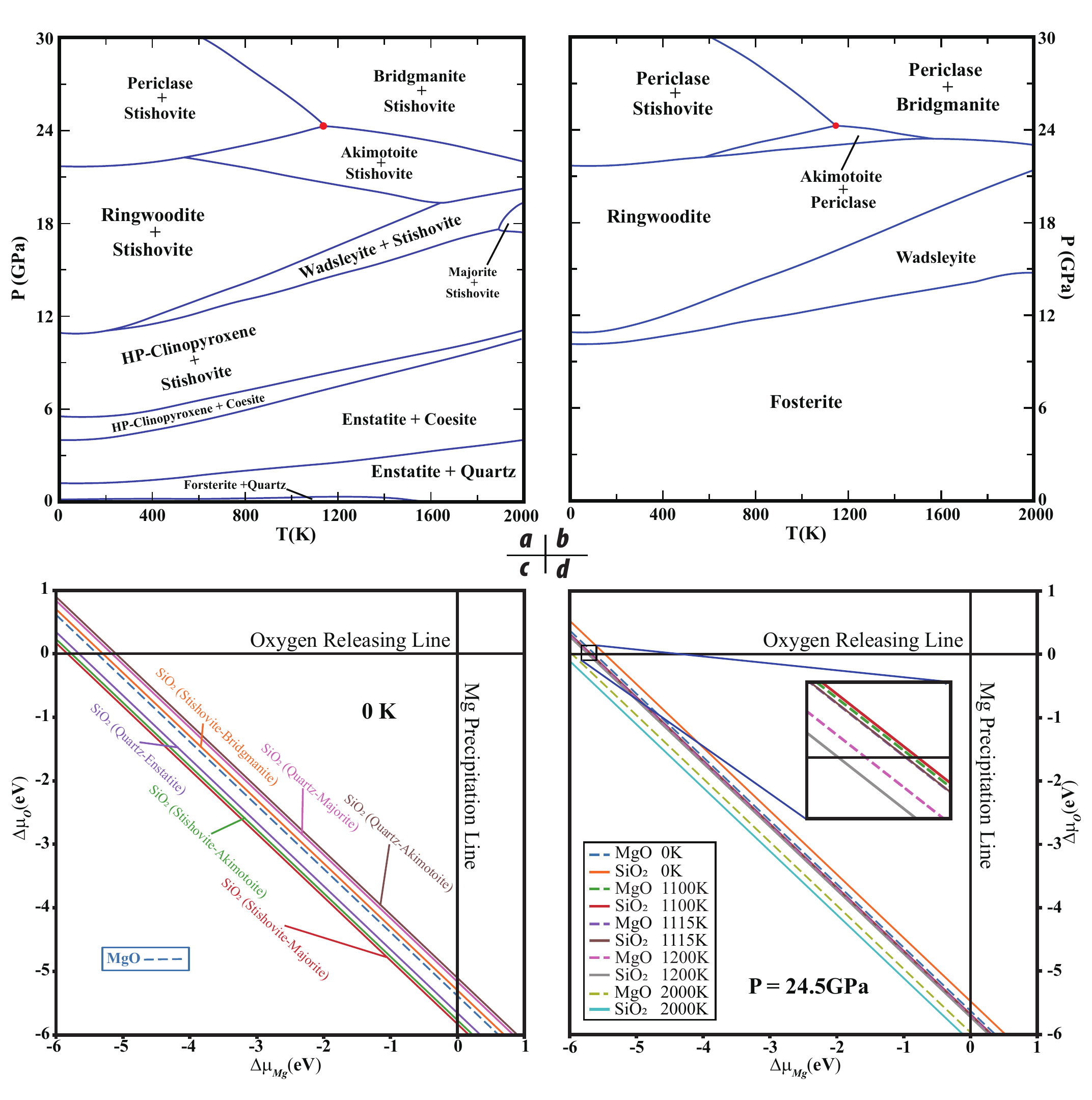}
 \caption{Phase Diagrams of semi-\ce{MgSiO3} systems, a) P-T diagram of \ce{MgSiO3}+\ce{SiO2} system, b) P-T diagram of \ce{MgSiO3}+\ce{MgO} system, c) $\Delta \mu $ diagram of different mineral combination, d) $\Delta\mu$ diagram of bridgmanite-stishovite in different temperature; The round dot in a) and b) is the transition point of bridgmanite break down to periclase+stishovite and the akimotoite transition to bridgmanite, it is around 24.5GPa and 1140 K.}
 \label{fig:PhaseD}
\end{figure}

The composition phase diagram 
based on experimentally determined thermodynamic data\cite{RN298417} 
is shown in Fig \ref{fig:PhaseD} a and b.
The precipitation criteria listed above for the various combinations of \ce{SiO2} and \ce{MgSiO3} are shown in Fig \ref{fig:PhaseD} c and d. 
In the $\Delta \mu$ graph, \ce{MgO} precipitates at the top left of the \ce{MgO} line, and \ce{SiO2} precipitates below the \ce{SiO2} line. 
The bridgmanite \ce{MgSiO3} phase is stable below the \ce{MgO} line and above the \ce{SiO2} line. 
The mineral 
is
stable only when the \ce{SiO2} line lies below on the left. 
Quartz is can coexist with enstatite and stishovite can coexist with akimotoite and majorite. 
Bridgmanite is not stable 
according to the DFT calculation at zero temperature,
but by including the vibrational contributions at finite temperature, bridgmanite becomes stable when the temperature is above 1115K and pressure is above 24.5GPa.
This is in good agreement 
with the
bridgmanite break down into periclase and stishovite in the P-T the phase diagram obtained from experimentally determined thermodynamic data. 
The region of stability of bridgmanite becomes larger when the temperature is increased, which is also consistent with the P-T phase diagram. 



\begin{figure}[h]
	\centering
	\includegraphics[width=0.9\linewidth]{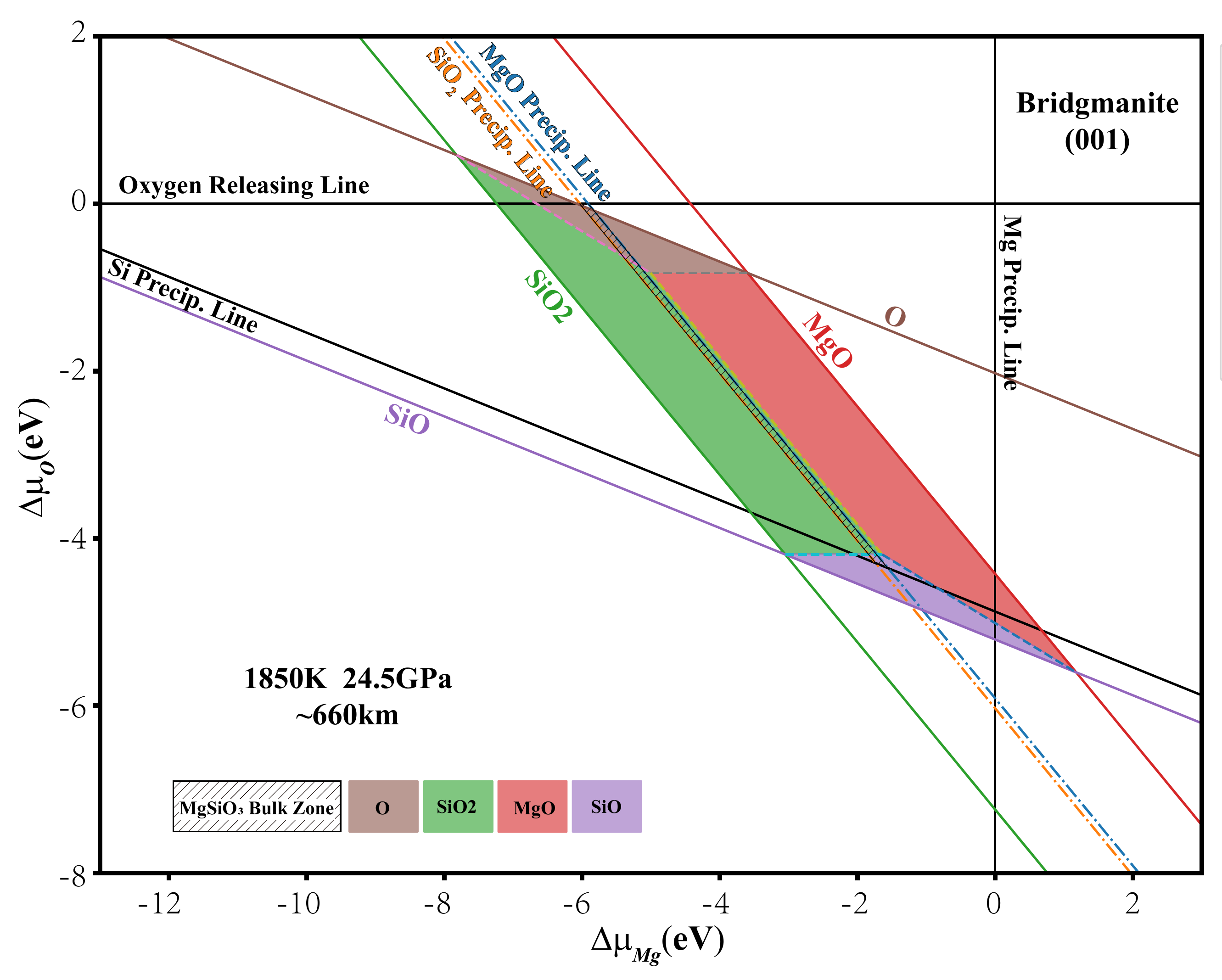}
	\caption{Phase Diagram of bridgmanite (001) surfaces at 24.5GPa and 1850K}
	\label{fig:001}
\end{figure}

The surface Gibbs free energy must be positive in order for the crystal to be stable

Fig. \ref{fig:001} shows the surface Gibbs free energy of four possible (001) surface terminations. Precipitation of silicon occurs below the Si precipitation line. Magnesium metal precipitates on the right of the Mg precipitation line. The region where a pure \ce{MgSiO3} surface can be obtained is the narrow stripe between the \ce{MgO} precipitation line on the right and the \ce{SiO2} precipitation line on the left. The lines marked with surface names are the boundaries where surface Gibbs free energies equals to zero. The surface Gibbs free energy is positive at the opposite side to the label. The names are marked at the negative side of the boundaries, which means that a \ce{MgSiO3} crystal will not be stable with respect to spontaneous surface formation beyond these boundaries. 

The surface Gibbs free energy of the (001) surfaces was calculated for 24.5GPa and 1850K which corresponds to about 660 km depth follow the geothermal gradient\cite{RN298416} in the deep Earth. The terminations with O and MgO are cataloged as Mg-center terminations, and SiO2 and SiO as Si-center termination.
Si-center terminations are more stable than Mg-center terminations in relatively reduced (i.e. oxygen-poor) environments. 
The Mg-center terminations become more stable than the Si-center terminations when the condition change from Mg-rich to Mg-poor.

\begin{figure}[h]
	\centering
	\includegraphics[width=0.9\linewidth]{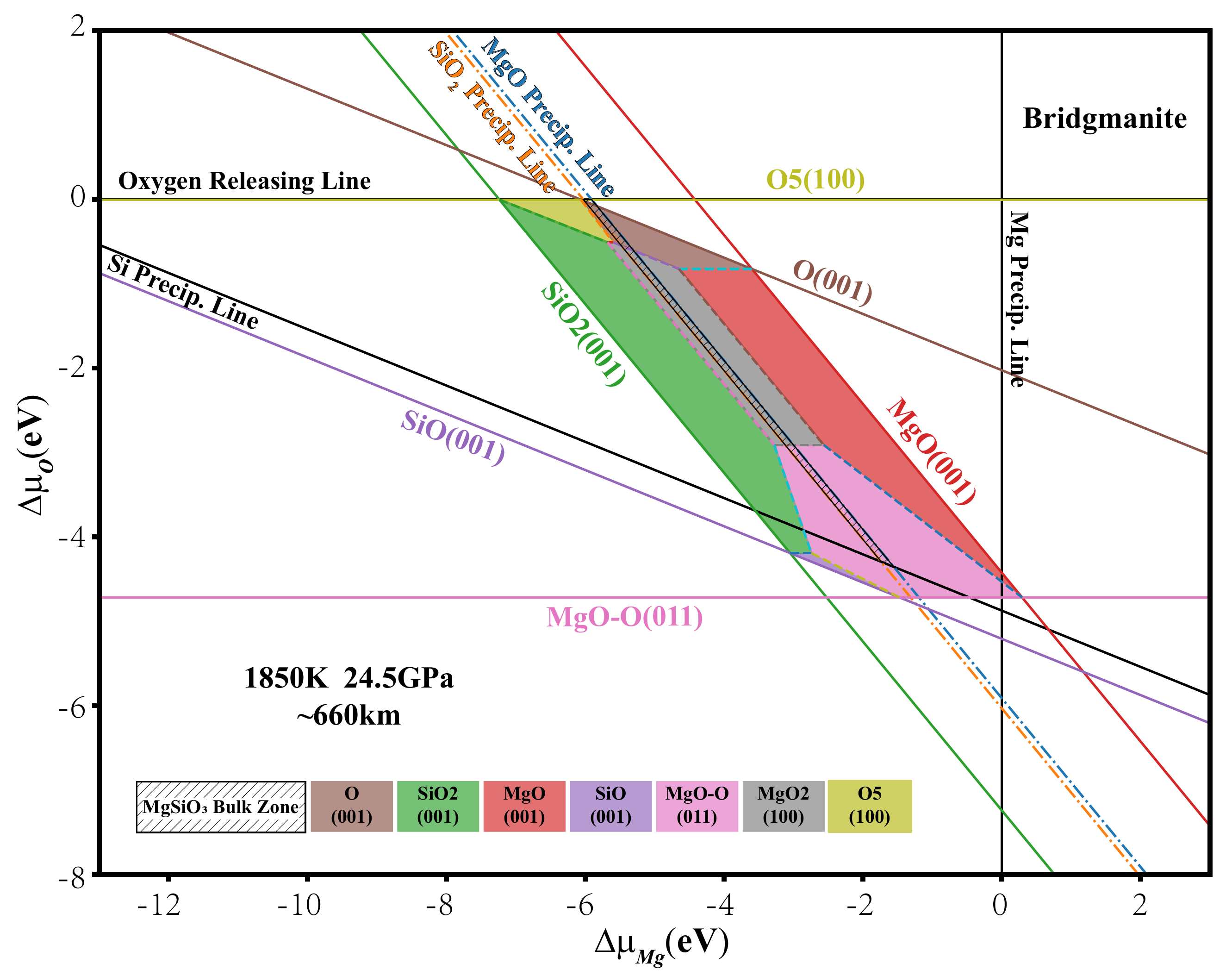}
	\caption{Phase Diagram of most stable bridgmanite surfaces among (001),(010),(100),(011) at 24.5GPa and 1850K}
	\label{fig:All_1850}
\end{figure}

The surface Gibbs free energy of all possible surface terminations of the (010),(100) and (011) surfaces was also calculated for 24.5GPa and 1850K (see Fig. \ref{fig:All_1850}). The most stable surface terminations are the same as for the (001) surface over most of the $\Delta \mu$ area. 
An oxygen rich termination, O5(100), occupies a small conner just below the oxygen releasing line. A MgO-O(011) termination becomes most stable in oxygen poor and magnesium rich conditions. A narrow belt lies in between the stable zone of MgO2(100). The results show that the (001) surfaces are most stable surface over a wide range of variation in Mg and O content. In the striped \ce{MgSiO3} stability zone, the most stable surface changes from O(001) to MgO2(100) and then MgO-O(011) at oxygen rich, medium and poor conditions respectively. 

It is likely that the dominant slip planes are these three most stable surface terminations under the corresponding reduction conditions in the deep Earth interior. 
Studies of the Earth mantle dynamics have been interpreted in terms of different slip planes \cite{RN298356,RN298354}. 
Our results indicate different explanations of the seismic anisotropy of the mantle and this will be pursued in future calculations.



In summary,
density functional theory based thermodynamic calculations have been carried out to analyze the relative stability of bridgmanite and the possible (001), (010), (100) and (011) surface terminations as a function of chemical potential of oxygen and magnesium.
The (001) surfaces are predicted to be the most stable under a wide range of conditions. Silicon rich terminations are more stable than magnesium based terminations under reducing (oxygen poor) conditions. Our results agree well with the experimentally determined thermodynamic data. 
Our results show that elevated temperature included through vibrational contributions plays an important role in stabilizing the bridgmanite crystal. 

This work was supported by National Science Foundation of China (Grants \#41503060 and \#41590620), 
Strategic Priority Research Program (B) of Chinese Academy of Sciences (\#XDB18000000 and \#XDB10020301). Computations were performed on resources provided by the Computer Simulation Lab, IGGCAS and the Icelandic High Performance Computing Center at the University of Iceland.

\section*{Conflicts of interest}
There are no conflicts to declare.




\renewcommand\refname{References}

\scriptsize{
\bibliography{rsc} 
\bibliographystyle{rsc} } 

\end{document}